\documentclass[aps,prd,secnumarabic,amssymb, amsmath,nobibnotes,nofootinbib,11pt]{revtex4} 
\usepackage{amsfonts,amsmath,hyperref,url, color}
\usepackage{bm, bbm}
\usepackage{graphicx}
\usepackage{mathtools}

\newcommand{\be}{\begin{equation}}
\newcommand{\bal}{\begin{align}}
\newcommand{\eal}{\end{align}}
\newcommand{\ee}{\end{equation}}
\newcommand{\bea}{\begin{eqnarray}}
\newcommand{\eea}{\end{eqnarray}}
\newcommand{\bit}{\begin{itemize}}
\newcommand{\eit}{\end{itemize}}

\newcommand{\ba}{\begin{aligned}}
\newcommand{\ea}{\end{aligned}}

\usepackage{color}

\begin{document}

\title{Infrared gravity and a celestial obstruction to monogamy constraints}

\author{Francesco Alessio}
\email{francesco.alessio@su.se}
\affiliation{\it \small NORDITA, KTH Royal Institute of Technology and Stockholm University, \\
 Hannes Alfv{\'{e}}ns v{\"{a}}g 12, SE-11419 Stockholm, Sweden\\}
\affiliation{\it \small  Department of Physics and Astronomy, Uppsala University,\\ Box 516, SE-75120 Uppsala, Sweden}

\author{Michele Arzano}
\email{michele.arzano@na.infn.it (corresponding author)}
\affiliation{Dipartimento di Fisica ``E. Pancini", Universit\`a di Napoli Federico II, I-80125 Napoli, Italy\\}
\affiliation{INFN, Sezione di Napoli,\\ Complesso Universitario di Monte S. Angelo,\\
Via Cintia Edificio 6, 80126 Napoli, Italy}

\begin{abstract}
We argue that gravitational interactions between particles require a departure from the conventional picture of the quantum state of a multiparticle system in terms of tensor products of one-particle states. This modification is essential in order to accommodate the existence of a new boost-like relativistic angular momentum charge which pairs of particles must carry asymptotically due to long-range effects of gravity. These findings challenge conventional assumptions, prompting a reevaluation of the constraints on quantum entanglement between particle subsystems in a black hole geometry.
\medskip
\begin{center}
{\small {\it Honorable Mention in the 2024 Gravity Research Foundation Essay Competition}}\\
\end{center}

\end{abstract}

\maketitle

In tackling the puzzling aspects of the quantum evolution of a black hole, one conundrum stands out as particularly perplexing: the need, in order to have a unitary and smooth evolution, for the degrees of freedom of the quantum field near the horizon to be entangled both with the ones in the interior of the black hole and with those far away from the horizon. This contradicts a property known as {\it monogamy of entanglement} \cite{Mathur:2009hf, Almheiri:2012rt} (see also \cite{Raju:2020smc} for a recent review).

The key assumption that underlies the conclusions leading to what has been dubbed ``monogamy paradox" is that the Hilbert space describing the degrees of freedom of a quantum field outside the black hole horizon can be factorized into two distinct subspaces one for modes near the horizon and one for those sufficiently far away from the latter.

The possibility of such factorization can be questioned invoking known problems in defining gauge invariant local observables in (quantum) gravity \cite{Marolf:2015jha, Donnelly:2016rvo} or resorting to arguments based on holographic properties of the quantum gravitational degrees of freedom \cite{Laddha:2020kvp}.

In this essay, we demonstrate that already ordinary, classical gravitational effects pose severe obstacles to the Hilbert space factorization assumed in the formulation of the monogamy paradox. Specifically, we show how classical leading-order self-force effects in gravitational scattering require the existence of a new {\it pairwise} quantum number between the modes of the field effectively providing an infrared obstruction to the factorization in sub-systems outside the black hole horizon.

Let us consider the scattering of point-particles mediated by the gravitational interaction. Their asymptotic trajectories, at late and early proper times $\tau_a$, exhibit the following logarithmic deviations from straight-line motion at first order in the Newton's constant $G$ \cite{Sahoo:2018lxl}
\begin{align}
\label{gr1}
&x_a^{\mu}(\tau_a)= \frac{p_a^{\mu}}{m_a}\tau_a\pm\log|\tau_a|\sum_{b\neq a}Gm_b\frac{u_a^{\mu}+ u_b^{\mu}\gamma_{ab}(2\gamma_{ab}^2-3)}{(\gamma_{ab}^2-1)^{\frac{3}{2}}}+\dots,\\&
p_a^{\mu}(\tau_a)= p_a^{\mu}\pm\frac{1}{\tau_a}\sum_{b\neq a}G m_b\frac{u_a^{\mu}+u_b^{\mu}\gamma_{ab} (2\gamma_{ab}^2-3)}{(\gamma_{ab}^2-1)^{\frac{3}{2}}}+\dots,
\end{align}
where the sums run over the total number of particles involved in the scattering, $u_a=p_a/m_a$ are their initial four-velocities, $\gamma_{ab}=-u_a\cdot u_b$ are the Lorentz factors and $\pm$ stays for late/early times.

If we look at the total relativistic angular momentum of the system, 
\begin{align}
\label{F1}
M^{\mu\nu}=\sum_{a}M^{\mu\nu}_{a}=\sum_{a}p_a^{\mu}x^{\nu}_{a}-p_a^{\nu}x^{\mu}_{a},
\end{align}
where $p_a^{\mu}=(E_a,p_a^i)$ and $x_a^{\mu}=(t,x^i)$ are the momenta and the positions of the particles involved in the scattering process,
we see that that there is an asymptotically non-vanishing contribution given by
\begin{align}
 \Delta M^{\mu\nu}=4G\sum_{a}\sum_{b\neq a}\frac{m_am_bu_a^{[\mu}u_b^{\nu]}\gamma_{ab}(2\gamma_{ab}^2-3)}{(\gamma_{ab}^2-1)^{\frac{3}{2}}}\log|\tau_a|+\dots\,,
\end{align}
which, written in the center of energy-momentum frame, reveals its boost-like nature. Indeed, considering for simplicity the case of a two-particle scattering we get \cite{Alessio:2024crv}
\begin{align}
\label{grsc}
\Delta N^z=\frac{2Gm_1m_2\gamma(3-2\gamma^2)}{\gamma^2-1}\log\left|\frac{\tau_1}{\tau_2}\right|+\dots.
\end{align}
Thus in the scattering process the system exchanges a boost charge or {\it mass moment}. 
The shift \eqref{grsc} of relativistic angular momentum during the scattering process has been referred to as {\it scoot} and it has been derived also in the relatively simpler case of electrodynamics \cite{Gralla:2021eoi,Gralla:2021qaf}. This effect indicates that asymptotic multiparticle states carry a non-vanishing charge which is {\it pairwise} i.e. it depends on the product of the masses of the two-particles and it is not present if one focuses on isolated particles.

The existence of such pairwise charge has dramatic consequences for the way we describe a collection of particles in relativistic quantum mechanics, as we now illustrate. 

In the quantum realm individual elementary particles are irreducible representations of the Poincaré group. Their spin and helicity have origin in the transformation of their states under the action of Wigner's little group: the subgroup of Lorentz transformations that leaves invariant a reference four-momentum \cite{Weinberg:1995mt}. These transformations act only on the particle's ``internal" quantum numbers.

Historically the first instance of an asymptotically non-vanishing {\it pairwise} charge was encountered in the study of the electromagnetic scattering of particles carrying both electric and magnetic charges \cite{Zwanziger:1972sx}. In this case the asymptotic angular momentum charge is related to a pairwise quantum number appearing in the transformation properties of a two-particle system under the action of a {\it pairwise little group}: the subgroup of Lorentz transformations which leaves {\it both} particles four-momenta unchanged. The pairwise little group in this case is easily identified as the group of rotations around one axis of motion \cite{Csaki:2020inw}, since one can always find a Lorentz transformation to a frame in which the pair is travelling along a given axis. This group is structurally the same as the little group accounting for the helicity of massless particles. The authors of \cite{Csaki:2020inw} dubbed the new quantum number ``pairwise helicity''. 

The gravitational scoot effect suggests that a pairwise quantum number should be present, not only in the rather exotic case of a system of interacting electric charges and magnetic poles, but 
{\it for any collection of gravitationally interacting massive particles}. 
The new pairwise relativistic angular momentum carried by multiparticle states is a boost charge and thus should be associated to the action of boosts on a two-particle system. It is obvious, however, that a boost acting on an ordinary two-particle configuration {\it does not} leave their four-momenta unchanged and thus it will not belong to the little group of the configuration which is supposed to act only on internal, pairwise quantum numbers!

The way out of this apparent contradiction is quite subtle and relies on the possibility of representing on-shell one-particle states in terms of ``celestial" quantum numbers. In this representation each state is denoted by a ket $|w,\bar{w}, \Delta, \sigma\rangle$ where $w$ and $\bar{w}$ are stereographic coordinates on the conformal boundary of the mass-hyperboloid and $\Delta$ and $\sigma$ are a boost and spin charge respectively: the latter is just the ordinary spin of the particle while the former is known as the conformal dimension of the state \cite{Pasterski:2016qvg,Pasterski:2017kqt}.
These states, known as conformal primary basis, transform under the unitary action of the Lorentz group as\footnote{We use the shorthand notation $|w,\bar w,\Delta,\sigma\rangle\equiv |w,\Delta,\sigma\rangle$.}
\begin{align}
    \label{F21}
U(\Lambda)|w,\Delta, \sigma\rangle=|cw+d|^{-2\Delta} 
\left(\frac{cw+d}{\bar{c}\bar{w}+\bar{d}}\right)^{\sigma}
|\Lambda w,\Delta, \sigma\rangle\,,
\end{align}
where $\Lambda w$ is a M\"obius transformation parametrized by complex numbers $a,b,c,d$ given by
\begin{align}
\Lambda w\equiv\frac{aw+b}{cw+d},\qquad ad-bc=1.
\end{align}
The internal quantum numbers $\sigma$ and $\Delta$ are related to the action of the little group which preserves the stereographic direction $w=0$, rather than the full reference four-momentum, and are eigenvalues of the generator of rotations around and boosts along a given reference direction  
\begin{equation}
    J_3 |0,\Delta, \sigma\rangle = -i\sigma |0,\Delta, \sigma\rangle\,,\qquad K_3 |0,\Delta, \sigma\rangle = -\Delta |0,\Delta, \sigma\rangle\,.
\end{equation}
We thus see that in this context the usual Wigner's little group for one-particle states is enlarged to include boosts in a given stereographic direction. 

For two-particle states one can always choose  a frame in which the stereographic boundary coordinate of the momentum of the first particle is $w_1 = 0$ and that of the second is $w_2 = \infty$ \cite{Lippstreu:2021avq}. If the two particles were massless this would correspond to the standard center of mass frame where their spatial momenta are aligned along the third axis.

The little group for the combined states $|0,\Delta_1, \sigma_1\rangle$ and $|\infty,\Delta_2, \sigma_2\rangle$ is non-trivial and it is again comprised of rotations and boosts which preserve both stereographic directions. Thus two-particle states must carry, besides the particles individual internal quantum numbers $\Delta_1, \sigma_1$ and $\Delta_2, \sigma_2$, two additional pairwise quantum numbers which we denote $\sigma_{12}$ and $\Delta_{12}$ so that
\begin{align}
    \label{mps}
&J_3 |0,\Delta_1;\infty,\Delta_2;\Delta_{12},\sigma_{12}\rangle= -i(\sigma_1+\sigma_2\pm\sigma_{12})|0,\Delta_1;\infty,\Delta_2;\Delta_{12},\sigma_{12}\rangle,\\\label{boost}
&K_3|0,\Delta_1;\infty,\Delta_2;\Delta_{12},\sigma_{12}\rangle=-(\Delta_1+\Delta_2\pm\Delta_{12})|0,\Delta_1;\infty,\Delta_2;\Delta_{12},\sigma_{12}\rangle,
\end{align}
where again $\pm$ stays for outgoing/incoming. Although the presence of a non-vanishing $\Delta_{12}$ emerges clearly in the celestial basis because it diagonalizes the action of boosts, it is also possible to observe a non-vanishing $\Delta_{12}$ in the standard on-shell momentum basis, as discussed in detail in \cite{Alessio:2024crv}.\\
The new pairwise quantum numbers $\sigma_{12}$ and $\Delta_{12}$ account, respectively, for the pairwise helicity \cite{Csaki:2020inw, Csaki:2020yei,Lippstreu:2021avq} and for the pairwise boost charge exchanged in the gravitational (and electromagnetic) scoot \cite{Alessio:2024wmz}. A general two-particle state $|w_1,\Delta_1, \sigma_1 ;w_2,\Delta_2,\sigma_2;\Delta_{12},\sigma_{12}\rangle$ will be thus labelled, not only by the quantum numbers of the individual particles, but also by the pairwise charges it carries. The existence of these additional degrees of freedom makes it clear that a two-particle state can no longer be described by a simple tensor product of the individual one-particle states. This fact was first pointed out in \cite{Zwanziger:1972sx} in the case of electromagnetic scattering of poles and charges in the standard on-shell momentum parametrization of one-particle states. Recently, in \cite{Csaki:2020yei}, it was recognized that the inclusion of a pairwise helicity quantum number  can be understood in terms of a non-trivial extension of multiparticle representations of the Poincaré group in which there exists an additional Hilbert space associated to the pairwise degrees of freedom.

As explained above, and as pointed out in our recent work \cite{Alessio:2024wmz}, the basis of conformal primary states allows to put pairwise helicity and the pairwise boost charge on the same footing: they are both related to the little group which leaves invariant two reference particle configurations in such representation. Generalizing the arguments of \cite{Csaki:2020yei} to conformal primary states we can conclude that a general two-particle state can not be factorized in the product of two one-particle states but rather must have the form
\begin{align}
    |w_1,\Delta_1, \sigma_1 ;w_2,\Delta_2,\sigma_2;\Delta_{12},\sigma_{12}\rangle = |w_1,\Delta_1, \sigma_1\rangle \otimes |w_2,\Delta_2, \sigma_2\rangle \otimes |\{w_1.w_2,\Delta_{12},\sigma_{12}\}\rangle\,. 
\end{align}
In this expression the first two terms are ordinary one-particle states in the conformal primary representation. The last factor $|\{w_1.w_2,\Delta_{12},\sigma_{12}\}\rangle$ carries the pairwise degrees of freedom. In the same way the states $|w_1,\Delta_1, \sigma_1\rangle \in \mathcal{H}_1$ and $|w_2,\Delta_2, \sigma_2\rangle \in \mathcal{H}_2$ are elements the one particle Hilbert spaces $\mathcal{H}_1$ and $\mathcal{H}_2$ represented in terms of celestial quantum numbers, the state 
$|\{w_1.w_2,\Delta_{12},\sigma_{12}\}\rangle \in \mathcal{H}_{12}$ carries the celestial pairwise quantum numbers belonging to both particles labelling states in a new pairwise Hilbert space $\mathcal{H}_{12}$.

Let us now go back to the gravitational scoot. Matching equation \eqref{boost} to \eqref{grsc} yields
\begin{align}
\label{match}
\Delta_{12}=\frac{2Gm_1m_2\gamma(3-2\gamma^2)}{\gamma^2-1}\log\bigg|\frac{\tau_1}{\tau_2}\bigg|.
\end{align}
Notice, crucially, that $\Delta_{12}\sim G$ and hence \textit{whenever gravity is switched on}, $\Delta_{12}$ cannot vanish. Such effect does not depend on the details of the scattering process under consideration and it is a consequence of the long-range nature of the gravitational interaction which alters the asymptotic behaviour of the particle trajectories \cite{Gralla:2021eoi,Gralla:2021qaf,Alessio:2024crv}. The infrared origin of the effect is corroborated by noticing that the scoot formula \eqref{grsc} bears similarities with the classical gravitational subleading soft factor discussed in \cite{Sahoo:2018lxl} and, indeed, the authors of \cite{DiVecchia:2022owy} conjectured that leading order in $G$ corrections to the particles trajectories could be inferred from the subleading soft graviton theorem \cite{Cachazo:2014fwa,Bern:2014vva,Bern:2014oka}.

As we argued above these infrared effects reflect in the appearance of a new pairwise quantum number which can only be captured by looking at the conformal boundary of the particles momentum space. 
Its presence requires the existence of a new Hilbert space 
inextricably linking the two particles when considered collectively as one system under the action gravitational self-interaction. The existence of this pairwise Hilbert space poses an obstruction to the description of a two-particle system as composed of two individual, separable, entities. Given the universal nature of gravity, such obstacle to factorization will be present regardless of the particle species and will apply to any collection of particles when described as modes of a quantum field. 

This brings us back to quantum fields in the presence of the horizon of a black hole and the trouble with their unitary evolution. The usual arguments resulting in conclusions that appear to reach a logical impasse assume that gravitational backreaction on the field modes can be safely neglected and that the multiparticle space of the theory is customarily described by a Hilbert space built from tensor products of individual one-particle states. We provided evidence that, as soon as gravitational interactions are turned on, leading order backreaction effects render this multi-particle space picture untenable. The existence of new pairwise degrees of freedom not only challenges the possibility of separating the fields quanta in distinct subsystems but also unveils an unexpected infrared escape route from monogamy paradoxes.



\subsection*{Acknowledgements}
The research of FA is fully supported by the Knut and Alice Wallenberg Foundation under grant KAW 2018.0116. MA acknowledges support from the INFN Iniziativa Specifica QUAGRAP. The research of MA was also carried out in the frame of Programme STAR Plus, financially supported by the University of Napoli Federico II and Compagnia di San Paolo.

\bibliography{bibliography}

\end{document}